# The effects of cap layer thickness on the performance of InGaN/GaN MQW solar cell

Saina Haghkish[1], Asghar Asgari[1, 2, *]

[1] Research Institute for Applied Physics and Astronomy, University of Tabriz, Tabriz 51665-163, Iran
[2] School of Electrical, Electronic and Computer Engineering, University of Western Australia, Crawley, WA 6009, Australia

Corresponding Author: asgari@tabrizu.ac.ir

Following letter introduces a theoretical approach to investigate the effect of two-step GaN barrier layer growth methodology on the performance of InGaN/GaN MQW solar cell, in which a lower temperature GaN cap layer was grown on top of each quantum well followed by a higher temperature GaN barrier layer. Different growth conditions would cause changes in the concentration of trap level density of states and imperfection sites. The simulation and comparison of 3 samples each with different cap layer thickness, reveals the fact that increasing cap layer thickness results in higher quantum efficiency, improved short circuit density of current and 3.2% increase of the fill factor.

The favorable physical properties of III-nitride semiconductors have recently attracted considerable attention for optoelectronics particularly for photovoltaic applications. InGaN possesses the most unique property of direct band gap ranging from 0.7ev for InN up to 3.4ev for GaN; therefor it covers most of the solar energy spectrum[1]. Moreover, InGaN takes advantages of high absorption coefficient [2](~ $10^5 \ cm^{-1}$), high radiation resistance[3] and the ability of incorporation with GaAs-based multi-junction solar cell[4], so it is the best candidate to design high efficiency MQW solar cells. On the other hand, growing InGaN films with enough thickness to absorb the entire incident solar radiation and relatively high indium content to reach optimal energy band gap, has proven to be difficult for many reasons[5]. First of all, InGaN layers are typically grown on GaN layers, so that, the more indium content incorporates, and the more lattice mismatch strain arises[6]. Additionally, to suppress formation of V-defects, it is common to arise the GaN barrier layer growth temperature and to use a mixture of hydrogen and nitrogen as the carrier gas[7,8,9,10]. These growth conditions can lead to Indium desorption from the InGaN QWs[8,11,12]. An innovative idea to preserve the high indium content inside the QWs while keeping the trap-level density of states below the critical amount is to perform a two-step GaN barrier layer growth method. Up to now this new method has been performed only through one experimental work[13], which reports improvement of the solar cell performance by increasing the thickness of low temperature GaN layer, called *cap layer*. The HAADF-STEM images and 3-D atom maps, reported by the mentioned experimental work has represented that the Indium fraction in QWs has been enhanced by increasing the cap layer thickness. Nevertheless there is a limit for the total thickness of the GaN region. An experiment has proved that the optimum thickness for the GaN layer between wells in a GaN/InGaN MQW solar cell is 6nm[14]. So by increasing the thickness of cap layer we have to decrease the barrier layer thickness to preserve the fixed thickness of 6 nm for the total GaN region. However the growth conditions for the barrier layer, consisting of the higher growth temperature and using a mixture of $H_2$ and $N_2$ as the carrier gas, makes it a region with less V-defect, traps or non-radiative recombination centers[15]. So decreasing the thickness of the barrier layer would cause a reduction of performance in the mentioned solar cell. Thus it is clear that finding the optimum thickness fraction for the cap and barrier layers seems necessary to optimize the total system efficiency. For the first time we have investigated cap layer thickness effect theoretically so we have performed a simulation for GaN/InGaN MQW solar cell structure, taking the advantage of adding a cap layer on top of each QW followed by growth of a higher temperature GaN barrier layer. Fig. 1 shows a cross-sectional schematic of the structure for modeled solar cell. This structure consists of a 2μm thick Si-doped n-GaN layer ([Si]=$6 \times 10^8 \ cm^{-3}$), a 10 period unintentionally doped InGaN/GaN MQW active region, a 30 nm Mg-doped p-GaN layer ([Mg]=$2 \times 10^{19} cm^{-3}$) and a degenerately doped 10 nm thick p⁺-GaN layer used for ohmic contacting. Wells and cap layers are both assumed to be grown in low temperature in a carrier gas consisting entirely of $N_2$, followed by growth of the high temperature GaN barrier layer with a mixture of $N_2$ and $H_2$ as the carrier gas. We will compare three samples which are varying in cap layer thickness from 1nm up to 3nm, and the relevant GaN barrier layers with thicknesses equal to 5nm, 4nm and 3nm for each sample, so that the overall thickness of the GaN regions between the InGaN QWs is fixed to 6nm.

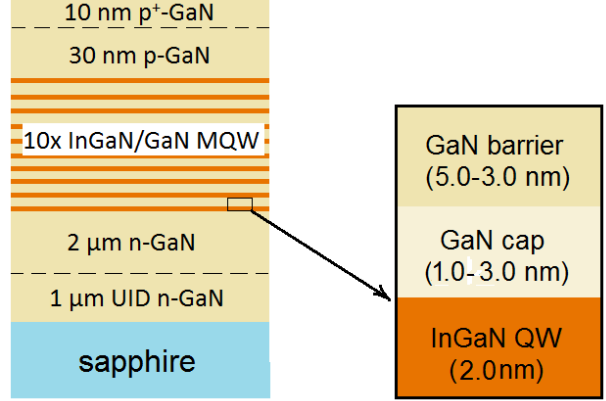

FIG. 1. The cross-sectional schematic of modeled GaN/InGaN MQW solar cell

To calculate the solar cell parameters first we have considered the theoretical basis of absorption in a QW[16]. Due to the Fermi Golden Rule and in absence of excitonic transitions we can write the absorption coefficient[16]:

$$\alpha(\hbar\omega_k) = \frac{e^2 \langle|P_{cv}|^2\rangle_{QW}}{\epsilon_0 m_0^2 c \hbar \eta \hbar \omega_k L} \mu \langle \phi_{en} | \phi_{hm} \rangle \ [f(E_f) - f(E_i)] \ H(E_{gmn} - \hbar\omega_k) \qquad (1)$$

$\langle|P_{cv}|^2\rangle_{QW}$ is the interaction probability which is equal to $|\langle u_c|\vec{r}.\vec{e}|u_v\rangle|^2$, $\mu$ is the reduced mass, $\phi_h(z)$ & $\phi_e(z)$ are envelope functions for holes and electrons, $f(E_i)$ & $f(E_f)$ are the Fermi occupational probabilities for charge carriers in the respective states and H is the Heaviside step function. Knowing α, we can evaluate the performance of the photovoltaic system by calculating the short-circuit current, open-circuit voltage and fill-factor which are all parameters associated with the current-voltage (J-V) diagram. Following equation gives the photo-generated current density in a MQW solar cell[17]:

$$J_G = \int_{\lambda_1}^{\lambda_2} q\phi(\lambda)(1 - R(\lambda)) \int_{x_j}^{x_j+w} e^{-\alpha(\lambda)x_j}(1 - e^{-\alpha(\lambda).w}) \qquad (2)$$

Where $\alpha(\lambda)$ is absorption coefficient, $R(\lambda)$ is reflectance and $\phi(\lambda)$ is the energy density of the incident light flux per unit area and per unit wavelength. $\lambda_1$ and $\lambda_2$ are the bottom and top of the range of absorption wavelength. It's quite clear that increasing the cap layer thickness which is associated with increasing Indium fraction in QWs, would decrease the band gap of the active region (InGaN) therefore the absorption edge moves to the lower energies. This shift of absorption edge could increase the photo-generated current for 2 reasons: first, the area with lower energy in the solar spectrum has higher radiation intensity and second, the red shift of absorption edge provides the wider interval for light absorption. Yet, the short-circuit current($J_{sc}$) is obtained by subtracting the recombination current ($J_R$) from the photo-generated current($J_G$). $J_R$ is a product of recombination of the photo-generated charge carriers within the intrinsic area and is calculated as[17]:

$$J_R = qwR_i = qwB_i n_i^2 \exp\left(\frac{qv}{kT}\right) \tag{3}$$

Where $R_i$ is the average recombination rate, $B_i$ is the recombination coefficient, $n_i$ is the charge carrier concentration and $v$ is the bias potential so that the index $i$ represents the intrinsic area of the p-i-n MQW solar cell. The intrinsic region of the studied solar cell, consists of InGaN and GaN as the wells, caps and barriers, each with unique physical and structural properties such as thickness, carrier concentration and recombination coefficient. However as mentioned earlier, the barrier layer, comparing to both other intrinsic layers, has fewer non-radiative recombination centers, due to its improved growth conditions. So by negligence of the recombination current in this layer one may write:

$$J_R = J_R^{cap} + J_R^{well} \tag{4}$$

$$J_R = qw_c n_c^2 B + qw_w n_w^2(x) B \tag{5}$$

$w_c$ is the overall thickness of cap layers, $w_w$ is the overall thickness of wells, $n_c$ is the carrier concentration in the cap layers and $n_w(x)$ is the carrier concentration in wells which is a function of Indium molar($x$). To expand Eq. (5), it is necessary to have the structural parameters for InGaN, so considering Eq. 6 and by inserting the values[18] for GaN and InN, one can extract the respective values for InGaN.

$$f^{In_xGa_{(1-x)}N} = x \times f^{InN} + (1-x) \times f^{GaN} + b \times x \times (1-x) \tag{6}$$

According to the Eq. 6, we can assume the carrier concentration of InGaN:

$$n^{In_xGa_{(1-x)}N} = n^{InN} \times (x) + n^{GaN} \times (1-x) = n^{GaN}(1 + \frac{n^{InN}}{n^{GaN}} x) \tag{7}$$

Considering that $n^{GaN} = n_c$ and $n^{In_xGa_{(1-x)}N} = n_w$, and:

$$n_w^2(x) = n_c^2 \times (1 + 2.5x)^2 \tag{8}$$

By applying the Eq. 8 in to Eq. 5 we obtain:

$$J_R = qBn_c^2[w_c + w_w(1+2.5x)^2] \tag{9}$$

Where $x$ is the Indium molar fraction. Now the short-circuit current could be obtained:

$$J_{sc} = J_G - J_R \tag{10}$$

However the open circuit voltage is a more complicated function of cap layer thickness and the other characteristics of the samples. One reason for this complicacy is that in addition of the built-in potential which is a function of layers' thickness, there is another potential available in our system, the piezoelectric induced voltage, which is a function of the Indium molar fraction in the QWs. Fig. 2 shows a cross-sectional image of a GaN/InGaN MQW solar cell with the space electric and induced electric charges. The mentioned charges are responsible for the present electric fields inside the structure, the potential barrier height and the open-circuit voltage[19].The spontaneous polarization ($P_{sp}$) of Ga-face InGaN layer in terms of Indium composition (x) is given by[20]: $P_{sp} = -0.003 \times x - 0.029$. Spontaneous polarization is the inherent property of the InGaN layer and produced due to non-centro-symmetry property of the wurtzite structure. The piezoelectric polarization in InGaN layer is developed due to lattice mismatch and different thermal expansion coefficient between GaN and InGaN layers. These piezoelectric polarization charges are distributed identically in the entire thickness of the InGaN layer for the well thickness less than 3 nm[19].

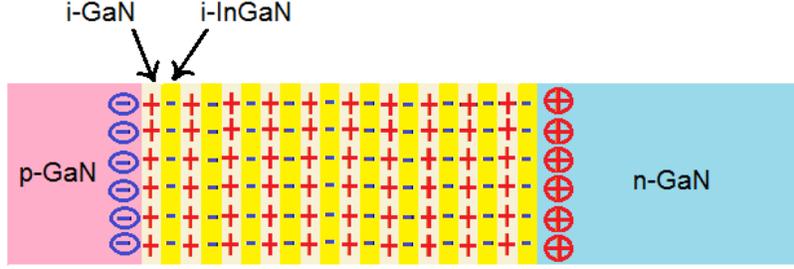

FIG. 2. Space charge and polarization induced charge distribution of a GaN/InGaN MQW solar cell[19]

The value of piezoelectric polarization ($P_{pz}$) in terms of Indium composition (x), is[21]: $P_{pz} = 0.176 \times x$. The net polarization is the sum of both spontaneous polarization and piezoelectric polarization in InGaN layer: $P = P_{sp} + P_{pz}$, so The net polarization electric field across each quantum well is[22]: $P = P_{sp} + P_{pz}$ and the net polarization electric field across each quantum well is[23]: $E_p = -(\frac{P}{\varepsilon})$, Where $\varepsilon$ is the permittivity of the InGaN layer and calculated by $\varepsilon = \varepsilon_0 \times \varepsilon_r$ where $\varepsilon_0$ is the vacuum permittivity and $\varepsilon_r$ is the relative permittivity and is calculated by $\varepsilon_r = 10.4 + (3.9 \times x)$ where $x$ is the indium molar fraction of InGaN layer[24]. The open circuit voltage ($V_{oc1}$) is the maximum voltage available across the solar cell for zero current and is given by[25]:

$$V_{oc1} = \frac{kT}{q} \times ln(\frac{J_{sc}}{J_0} + 1) \tag{11}$$

Where $K$ is the Boltzman constant, $T$ is the operating temperature of the solar cell, $q$ electric charge, $J_{sc}$ is the photo-generated current and $J_0$ is the reverse saturation current in the solar cell. The reverse saturation current in a p-n solar cell is given by[24] $J_0 = q[\frac{D_n \times (n_i^{p-GaN})^2}{l_n \times N_A^{p-GaN}} + \frac{D_p \times (n_i^{n-GaN})^2}{l_p \times N_D^{n-GaN}}]$ and it is sum of the reverse currents created by minority carriers of each region. Where $D_n$ and $D_p$ are the diffusion coefficients for minority carriers, the electrons in p-GaN and the holes in n-GaN. The $l_n$ and $l_p$ are the minority carrier diffusion lengths. In the case of a p-i-n solar cell we have to add the portion of intrinsic region minority carriers current to the mentioned Eq. Considering that $n_p^{p-GaN} = \frac{(n_i^{p-GaN})^2}{N_A^{p-GaN}}$ and $p_n^{n-GaN} = \frac{(n_i^{n-GaN})^2}{N_D^{n-GaN}}$, and the fact that un-doped GaN is n-type, one may write:

$$J_0 = q[\frac{D_n \times n_p^{p-GaN}}{l_n} + \frac{D_p \times p_n^{uid-GaN}}{l_p} + \frac{D_p \times p_n^{n-GaN}}{l_p}] \tag{12}$$

Where the index '$uid$' indicates unintentionally-doped GaN. Regarding the facts $p_n^{uid-GaN} \gg p_n^{n-GaN}$ and $n_p^{p-GaN}$, we can assume that the reverse saturation current is mainly a function of minority carrier concentration of the intrinsic region, so that altering intrinsic layers will change the $J_0$. Specifically increasing the cap layer thickness would arise the leakage current due to this layer's more trap level concentration. The built-in potential is given as [24] :

$$V_{bi} = \frac{kT}{q} \ln[\frac{N_A^{p-GaN} N_D^{n-GaN}}{n_i^{p-GaN} n_i^{n-GaN}}] \tag{13}$$

This is the maximum value of the open circuit voltage in the ideal case of the GaN p-n solar cell. The ratio of the practical value of open circuit voltage, $V_{oc1}$, to the ideal value of open-circuit voltage (built-in voltage), $V_{bi}$, in a solar cell gives the loss factor (F)[19].

$$F = V_{oc1}/V_{bi} \tag{14}$$

This loss factor, F, is the representation for all of phenomena's responsible for the carrier losses in the entire structure of the device which reduces the output voltage of the p-n solar cell. In the case of a MQW p-i-n solar cell, when the intrinsic MQW area is inserted into the depletion region of the p-n solar cell an additional voltage is developed in the depletion region due to piezoelectric field at every GaN/InGaN interface. Then the total polarization voltage developed in MQW is the total of all the voltages develop in each individual interface and calculated by[22]:

$$V_p = -E_p \times d_w \tag{15}$$

Where $d_w$ is the total thickness of quantum wells and calculated by $dw = d_1 + d_2 + d_3 + d_4$ where $d_1$, $d_2$, $d_3$ and $d_4$ are the individual thicknesses of each QW. The $E_p$ is the piezoelectric field in single QW. This polarization voltage ($V_p$) develops in the same region where the built-in voltage ($V_{bi}$) is developed. Thus polarization voltage will suffer from the same carrier losses which work for built-in voltage[19]. Then the value of open circuit voltage due to net polarization electric field is calculated by multiplying polarization voltage ($V_p$) with loss factor ($F$). So the effective piezoelectric induced voltage is given by[19]:

$$V_{oc2} = F \times V_p \tag{16}$$

Therefore the total open circuit voltage ($V_{oc}$) is sum of both parts of open circuit voltage developed due to built-in voltage and polarization induced voltage[19].

$$V_{oc} = V_{oc1} + V_{oc2} \tag{17}$$

Table 1 represents the calculated values for The photo-generated current density, the recombination current and the resultant short-circuit current. As shown in the table, the photo-generated current increased with increasing the cap layer thickness, because the shifting of absorption edge to lower energies means the absorption happens in the region of solar spectrum with higher photon intensity. The above results also confirm that the short-circuit current increases with increasing the cap layer thickness.

TABLE 1. The photo-generated, recombination, and $J_{SC}$ for the 3 samples with different cap thicknesses

| Cap | In fraction (%) | $J_G$ (mA/cm$^2$) | $J_R$(mA/cm$^2$) | $J_{sc}$(mA/cm$^2$) |
|---|---|---|---|---|
| 1nm | 0.09 | 0.8367 | 0.4877 | 0.3490 |
| 2nm | 0.15 | 1.3167 | 0.6670 | 0.6487 |
| 3nm | 0.21 | 1.8932 | 0.8279 | 1.0563 |

As it is discussed before, since increasing the cap layer thickness helps preserving more indium molar fraction in quantum wells, so the piezoelectric open-circuit voltage ($V_{oc2}$) through the entire solar cell will increase. $V_{oc2}$ is the only term of open-circuit voltage appearing in dark J-V curve, so the samples with thicker cap will have higher open-circuit voltage under darkness. Besides, the cap layer suffers from more crystalline imperfections due to its un-optimized growth conditions. Therefore increasing the cap layer thickness would arise the reverse saturation current($J_0$) and recombination current($J_R$), so $V_{oc1}$ would decrease. For the devices under illumination, These 2 processes act against each other. The optimum condition where the $V_{oc}$ reaches its maximum value happens for the sample with 2nm cap thickness (Fig. 3). At these circumstances the maximum fill-factor is also obtained as it is shown in the Table 2.

TABLE 2. The non-piezoelectric open-circuit voltage($V_{oc1}$) and loss factor($F$) for 3 samples with different cap thicknesses

| Cap | $V_{bi}$(volt) | $V_{oc1}$(volt) | F(%) | $V_p$(volt) | $V_{oc2}$(volt) | $V_{oc}$(volt) | FF (%) |
|---|---|---|---|---|---|---|---|
| 1nm | 2.56 | 0.95 | 0.372 | 3.97 | 1.479 | 2.43 | 59.0 |
| 2nm | 2.56 | 0.78 | 0.304 | 6.00 | 1.824 | 2.60 | 63.2 |
| 3nm | 2.56 | 0.67 | 0.260 | 6.99 | 1.822 | 2.49 | 61.1 |

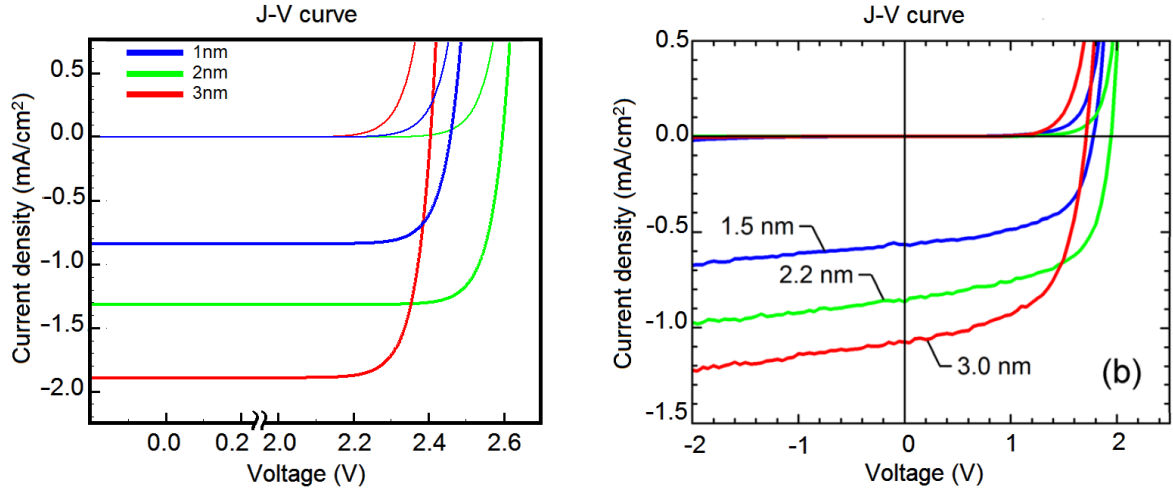

FIG.3. (a) simulated diagram for dark and illuminated J-V curves from samples with varying GaN cap layer thickness, (b) the experimental J-V curves for similar samples[13]

In summary, we introduced a theoretical approach to investigate the effect of cap layer thickness on the performance of a GaN/InGaN MQW solar cell. After solving the Schrodinger equation for electrons-holes and obtaining the structure profile, we obtained the absorption and its resultant photo-current. We recognized two dominant factors determining the open-circuit voltage of a MQW solar cell: the built-in potential which is a function of layers' thickness and the piezoelectric induced voltage, which is a function of the Indium molar fraction in the QWs. The optimum thickness for cap layer is about 2nm where the solar cell reaches its best performance with the fill factor of 63.2%.